\documentclass[11pt]{article}
\usepackage[top=1in, bottom=1in, left=1in, right=1in]{geometry}
\usepackage{amsmath}
\usepackage{mathrsfs}
\usepackage{epsfig}
\usepackage{graphicx}
\usepackage{cite}
\usepackage{url}
\usepackage{authblk}
\usepackage{color}

\begin{document}

\renewcommand{\arraystretch}{1.7}

\title{Message-Passing Methods for Complex Contagions}

\date{16 Mar 2017}
\author{James P. Gleeson}
\affil{MACSI, Department of Mathematics and Statistics, University of Limerick, Ireland}
\author{Mason A. Porter}
\affil{Department of Mathematics, University of California, Los Angeles, California 90095, USA}

\maketitle
\begin{abstract}

Message-passing methods provide a powerful approach for calculating the expected size of cascades either on random networks (e.g., drawn from a configuration-model ensemble or its generalizations) asymptotically as the number $N$ of nodes becomes infinite or on specific finite-size networks.
We review the message-passing approach and show how to derive it for configuration-model networks using the methods of \cite{Gleeson07,Gleeson08,Dhar97}. Using this approach, we explain for such networks how to determine an analytical expression for a ``cascade condition'', which determines whether a global cascade will occur.
We extend this approach to the message-passing methods for specific finite-size networks, and we derive a generalized cascade condition. Throughout this chapter, we illustrate these ideas using the Watts threshold model.
\end{abstract}




\section{Introduction}

In this chapter, we consider analytical approaches for calculating the expected sizes of cascades in complex contagion.\footnote{See  \cite{PorterFrontiers} and references therein for discussions of cascades on networks and for the definition of a complex contagion. See \cite{motter2017} for a friendly introduction to cascades on networks.}
As a concrete example of complex contagion, we use the Watts threshold model (WTM) \cite{Watts02} (see also \cite{granovetter78,valente-book}) on undirected, unweighted networks. In this model, each node $i$ of a network has a positive threshold $r_i$; usually, the thresholds are chosen at random from a given probability distribution, but (with some difficulty and arguably circular reasoning) they can also be estimated from empirical data.
We focus in particular on the case in which a contagion is initiated by multiple seed nodes, so that
a finite (but small) fraction of the network nodes are assumed to be active at the beginning of contagion dynamics.

Each node can be in one of two states; we will call the states ``inactive'' and ``active''. All nodes, except for the seed nodes, are initially inactive. In each discrete time step, each inactive node $i$ of a network considers its neighboring nodes, and it becomes active if the fraction of its neighbors that are active exceeds (or equals) the threshold $r_i$ of node $i$.
One can interpret the threshold as a node's stubbornness and the fraction of active nodes as one choice of ``peer pressure'' function \cite{MelnikMulti}.
Once a node becomes active, it cannot later return to the inactive state, so the cascade grows in a monotonic fashion. An important macroscopic quantify is the fraction $\rho_n$ of active nodes at time step $n$. Because of the monotonic nature of the dynamics, $\rho_n$ is a nondecreasing function of $n$, so (because $\rho_n \leq 1$, by definition), the limit $\rho_\infty = \lim_{n\to \infty} \rho_n$ exists. We call $\rho_\infty$ the ``steady-state fraction of active nodes'', and we focus our attention on methods for analytically approximating its value\footnote{In \cite{GleesonSPIE}, Gleeson and Cahalane showed that if nodes are updated one at a time in a random order, rather than all simultaneously as described here (i.e., if we use ``asynchronous'' updating instead of ``synchronous'' updating \cite{PorterFrontiers}), one obtains the same steady-state limit $\rho_\infty$, although the temporal evolution of the active fraction does depend on the updating scheme that is used \cite{Fennell16}.
See Sec.~5.1 of \cite{PorterFrontiers} for a description of an algorithm for a stochastic simulation of the WTM.}.
Assuming that a fraction $\rho_0$ of the nodes are selected uniformly at random as the seed nodes for a contagion, we want to predict the steady-state value $\rho_\infty$ and to determine the conditions under which $\rho_\infty$ substantially exceeds $\rho_0$. In other words, we want to answer the question ``when does a global cascade occur?''.\footnote{See the discussion in \cite{PorterFrontiers} of ways of measuring cascade sizes.}


In Secs.~\ref{sec:GP2} and \ref{sec:GP3}, we focus on ensembles of infinite-size random networks (i.e., on asymptotic behavior as the number $N$ of nodes becomes infinite), both without (see Sec.~\ref{sec:GP2}) and with (see Sec.~\ref{sec:GP3}) degree--degree correlations. In Sec.~\ref{sec:GP4}, we discuss recent progress on calculating $\rho_\infty$ for finite-size networks. We conclude in Sec.~\ref{sec:GP5}.


\section{Configuration-model networks}\label{sec:GP2}

Let's begin by assuming that our networks are realizations drawn from a configuration-model ensemble \cite{Fosdick2016}; they are characterized by a given degree distribution $p_k$, where $p_k$ is the probability that a node chosen uniformly at random has $k$ neighbors, but they are otherwise maximally random. Moreover, our theoretical approach is for the limit of infinitely large networks (sometimes called the ``thermodynamic limit''). Because configuration model networks are locally tree-like \cite{Melnik11}, one might expect that we can apply mean-field approaches, such as those used for models of biological contagions \cite{Pastor-Satorras15},
to approximate the fraction of active nodes. We'll first briefly summarize what we'll call a ``naive mean-field (MF)'' approach, and we'll then explain why---and how---it can be improved.


\subsection{Naive mean-field approximation} \label{naive}

We define $\rho_n^{(k)}$ as the probability that a node of degree $k$ is active at time step $n$; the total fraction of active nodes is then given by
\begin{equation} \label{GP1}
	\rho_n = \sum_k p_k \rho_n^{(k)}\,.
\end{equation}
A node of degree $k$ is active at time $n$ either because (i) it was a seed node (with probability $\rho_0$), or (ii) it was not a seed node (with probability $1-\rho_0$), but it has become active by time step $n$. In the latter case, the number $m$ of its active neighbors at time $n-1$ must be large enough so that the fraction $m/k$ is at least as large as the node's threshold.
Treating the $k$ neighbors as independent of each other, the probability that $m$ of the $k$ are active at time $n$ is given by the binomial distribution
\begin{equation}
	{k\choose m}   (\overline{\rho}_{n-1})^m (1-\overline{\rho}_{n-1})^{k-m}\,,
\end{equation}
where $\overline{\rho}_{n-1}$ is the probability that the node at the end of a uniformly randomly chosen edge is active at time step $n-1$. Under the usual mean-field assumptions (see, for example, \cite{Melnik11}), we write $\overline{\rho}_{n-1}$ as the weighted mean over the possible degrees of neighbors\footnote{The weighting $k/z p_k$ arises because we are considering the mean over nodes of degree $k$, where those nodes are reached by traveling along an edge from the node of interest. It is well-known (see, e.g., \cite{Newmanbook}) that a node at the end of a uniformly randomly chosen edge of a configuration-model network
has degree $k$ with probability $k/z p_k$, reflecting the fact that high-$k$ nodes are more likely than low-$k$ nodes to be reached in this way.}
\begin{equation}
	\overline{\rho}_{n-1} = \sum_{k}\frac{k}{z} p_{k} \rho_{n-1}^{(k)}\,, \label{GPweightedavg}
\end{equation}
where $z=\sum_k k p_k$ is the mean degree of the network.

\begin{figure}
\centering
\vspace{-2cm}
\epsfig{figure=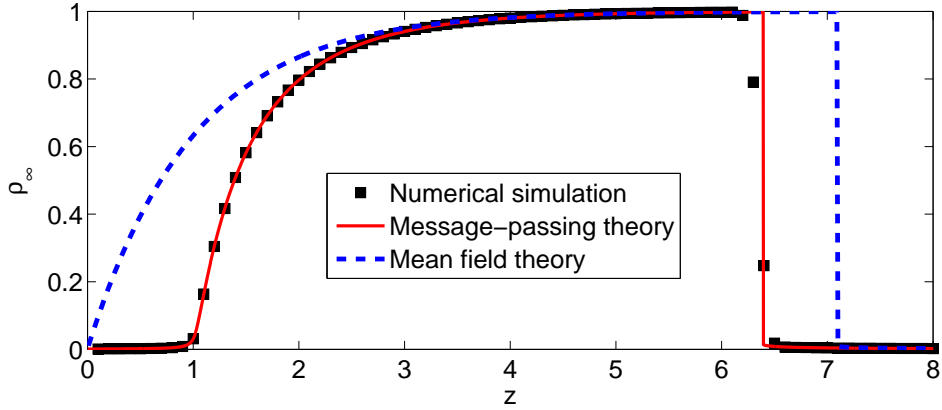,width=14.5 cm}
\vspace{-2.2cm}
\caption{The expected steady-state fraction $\rho_\infty$ of active nodes for cascades in the Watts threshold model (WTM), where every node has the same threshold $r=0.18$ (so a node becomes active when its fraction of active neighbors is at least as large as $18\%$).
The networks are Erd\H{o}s--R\'{e}nyi random graphs ($G(N,m)$, where $m$ is the total number of edges) with mean degree $z$ (so they have a Poisson degree distribution $p_k = z^k e^{-z}/k!$), and the initial seed fraction is $\rho_0=10^{-3}$.
The numerical simulation results, shown by the black squares, are a mean over 100 realizations on networks with $N=10^5$ nodes. The blue dashed curve shows the result of the naive mean-field approximation given by Eqs.~(\ref{GP4}) and (\ref{GP4a}), and the red solid curve is for the message-passing approach of Eqs.~(\ref{GP8}) and (\ref{GP10}).} \label{fig:GP1}
\end{figure}

If $m$ neighbors are active, the probability that the node is active is equal to the probability that its threshold is less than $m/k$.
We write this probability as $C\left(m/k\right)$, where $C$ is the cumulative distribution function (CDF) of the thresholds. Putting together these arguments and summing over all possible values of $m$, we write the MF approximation for $\rho_n^{(k)}$ as
\begin{equation} \label{GP3}
	\rho_n^{(k)} = \rho_0 + (1-\rho_0) \sum_{m=0}^k {k\choose m}   \left(\overline{\rho}_{n-1}\right)^m \left(1-\overline{\rho}_{n-1}\right)^{k-m} C\left(\frac{m}{k}\right)\,.
\end{equation}
Multiplying Eq.~(\ref{GP3}) by $\frac{k}{z}p_k$ and summing over $k$ gives
\begin{equation} \label{GP4}
	\overline{\rho}_n = \rho_0 + (1-\rho_0) \sum_k \frac{k}{z} p_k \sum_{m=0}^k {k\choose m}   \left(\overline{\rho}_{n-1}\right)^m \left(1-\overline{\rho}_{n-1}\right)^{k-m} C\left(\frac{m}{k}\right)\,,
\end{equation}
so we now have an expression for $\overline{\rho}_n$ in terms of $\overline{\rho}_{n-1}$. Starting from an initial condition with a the fraction $\overline{\rho}_0=\rho_0$ of seed nodes (chosen uniformly at random), one can iterate Eq.~(\ref{GP4}) to determine $\overline{\rho}_n$ for any later time step, and it converges to $\overline{\rho}_\infty$ as $n\to \infty$.
One then calculates the naive MF approximation to the steady-state fraction of active nodes $\rho_\infty$ from Eqs.~(\ref{GP1}) and (\ref{GP3}) with the formula
\begin{equation}
	\rho_{\infty} = \rho_0+ (1-\rho_0) \sum_k p_k \sum_{m=0}^k {k\choose m}  \left(\overline{\rho}_{\infty}\right)^m \left(1-\overline{\rho}_{\infty}\right)^{k-m} C\left(\frac{m}{k}\right)\,. \label{GP4a}
\end{equation}
 However, as we illustrate in Figure~\ref{fig:GP1}, the naive MF approximation calculated using Eqs.~(\ref{GP4}) and (\ref{GP4a}) does not accurately match the values of $\rho_\infty$ from numerical simulations on large networks. In Sec.~\ref{sec:config}, we consider why this mismatch occurs, and we introduce an improved approximation technique, which is of ``message-passing'' type.

\subsection{Message-passing for configuration-model networks}\label{sec:config}


In this section, we present the approach that was first used in \cite{Gleeson07,Gleeson08}, who adapted the method used by Dhar et al.~\cite{Dhar97} for the zero-temperature random-field Ising model on Bethe lattices. Nowadays, the approach is called ``message-passing for configuration-model networks''. See, for example, Sec.~IV of \cite{shrestha14}.

\begin{figure}
\centering
\vspace{1cm}
\epsfig{figure=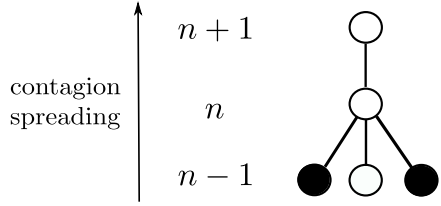,height=2.7cm}
\caption{Schematic for the method described in Sec.~\ref{sec:config}. We suppose that the contagion spreads upward from level $n-1$ to level $n$ and beyond. The assumption of infinite network size allows us to consider the limit of an infinite number of levels, terminating with the ``top'' (or ``root'') node of the tree approximation.} \label{fig:GP2}
\end{figure}

The fundamental problem with the naive MF approach of Sec.~\ref{naive} is that it neglects the directionality in the spreading of a contagion: the contagion spreads outwards from the seed nodes, and it can reach inactive nodes only after it has first infected some of their neighbors. In the schematic in Figure~\ref{fig:GP2}, we assume that the contagion spreads upward from ``level'' $n-1$ to level $n$ and then to level $n+1$. We number the levels according to their distance from the seed nodes, which we place at level $0$. This is a highly stylized approximation, as we are almost always considering networks that are not actually trees (and, e.g., social networks typically have significant clustering), but we see nevertheless that it gives good results (see, e.g., the discussion in \cite{Melnik11}). For the synchronous updating that we employ in this chapter, level $n$ of the tree approximation corresponds to time step $n$ of the contagion process on the original network. See \cite{Gleeson08} for details and extension to asynchronous updating.

We now focus again on the steady-state limit $n\to \infty$. We introduce the variable $q_n^{(k)}$, the probability that a node of degree $k$ on level $n$ is active, conditional on its parent (at level $n+1$) being inactive. When we calculate $q_n^{(k)}$, we account for the directionality of the contagion spreading because we assume that the node at level $n+1$ in Fig.~\ref{fig:GP2} is inactive at the time when the node at level $n$ is updating from the inactive to the (possibly) active state. As before, there are two ways in which the node at level $n$ can be active: either it was a seed node (with probability $\rho_0$), or it was not a seed node (with probability $1-\rho_0$) but has been activated by its children (i.e., the nodes at level $n-1$ in Fig.~\ref{fig:GP2}). Because the level-$n$ node has degree $k$ and one of its edges connects to its (inactive) parent, there are $k-1$ children node at level $n-1$. Each of these children is active with probability $q_{n-1}$, where (similar to Eq.~(\ref{GPweightedavg}) $q_n$ is the weighted mean over the $q_n^{(k)}$ values. That is,
\begin{equation} \label{GP5}
	q_{n-1} = \sum_{k}\frac{k}{z} p_{k} q_{n-1}^{(k)}\,.
\end{equation}
 Therefore, the probability that $m$ children are active is given by the binomial distribution on $k-1$ nodes, where each is active with  independent probability $q_{n-1}$. That is,
\begin{equation} \label{GP6}
	{{k-1}\choose m} (q_{n-1})^m (1- q_{n-1})^{k-1-m}\,.
\end{equation}
As with the naive MF case, the activation of a degree-$k$ node with $m$ active children depends on its threshold being less than the fraction $m/k$; this occurs with probability $C\left( m/k\right)$.
Putting together the preceding arguments, we write
\begin{equation} \label{GP7}
	q_n^{(k)} = \rho_0 + (1-\rho_0) \sum_{m=0}^{k-1} {{k-1}\choose m}   (q_{n-1})^m (1-q_{n-1})^{k-1-m} C\left(\frac{m}{k}\right)\,,
\end{equation}
and we obtain a discrete scalar map for $q_n$ by multiplying Eq.~\ref{GP7} by $\frac{k}{z} p_k$ and summing over $k$. Using Eq.~(\ref{GP5}) then yields
\begin{equation} \label{GP8}
	q_n = \rho_0 + (1-\rho_0)\sum_k \frac{k}{z} p_k \sum_{m=0}^{k-1} {{k-1}\choose m}   (q_{n-1})^m (1-q_{n-1})^{k-1-m} C\left(\frac{m}{k}\right)\,.
\end{equation}
Iterating Eq.~\ref{GP8} starting from initial condition $q_0=\rho_0$ leads to the steady-state value
\begin{equation} \label{GP9}
	q_\infty = \lim_{n\to \infty} q_n\,.
\end{equation}
Finally, we use the fact that a node at the ``top'' (or ``root'') of the tree---formally at level $\infty$---has $k$ children with probability $p_k$ and (assuming that the root node is not a seed node) that each child is active with probability $q_\infty$. We then determine the steady-state active fraction of nodes from $q_\infty$ by calculating
\begin{equation}
	\rho_\infty = \rho_0 + (1-\rho_0)\sum_k  p_k \sum_{m=0}^{k} {{k}\choose m}   (q_{\infty})^m (1-q_{\infty})^{k-m} C\left(\frac{m}{k}\right)\,. \label{GP10}
\end{equation}
The solid red curve in Figure~\ref{fig:GP1} shows the result of using Eqs.~(\ref{GP8}) and (\ref{GP10}) to determine the steady-state fraction of active nodes. Clearly, this approximation method is very accurate, and it is far superior to the naive MF approach of Sec.~\ref{naive}.


\subsection{The criticality condition (i.e., ``cascade condition'')}\label{sec:GPcc}

An additional benefit of the analytical approach that we outlined in Sec.~\ref{sec:config} is that it enables one to determine conditions on the model parameters that control whether or not global cascades occur. This question was first addressed by Watts \cite{Watts02} using a percolation argument, but one can derive same condition using the approach of Sec.~\ref{sec:config}. For this analysis, we assume that the seed fraction is vanishingly small, so we take the $\rho_0\to 0$ limit of our general equations. (See \cite{Gleeson07} for extensions to nonzero $\rho_0$.) In this case, Eq.~(\ref{GP8}) always has the solution $q_n\equiv 0$ for all $n$, corresponding to the case of no contagion. However, for certain parameter regimes, this
contagionless solution can be unstable, and then any infinitesimal seed fraction $\rho_0>0$ leads to a global cascade of nonzero fractional size. (The ``fractional size'' of a contagion is the number of active nodes divided by the total number of nodes.) Therefore, we linearize Eq.~(\ref{GP8}) about the solution $q_n\equiv 0$ to determine its (linear) stability. For scalar maps of the form $q_n=g(q_{n-1})$, the criterion for instability of the $0$ solution is that \cite{Strogatzbook}
\begin{equation} \label{GP11}
	|g'(0)| > 1\,.
\end{equation}
Differentiating the right-hand side of Eq.~(\ref{GP8}) and setting $q_{n-1} = 0$ yields the following condition for global cascades to occur (from infinitesimal seed fractions)\footnote{Note that $C(0)=0$, because we have assumed that all thresholds are positive.}:
\begin{equation} \label{GP12}
	\sum_k \frac{k}{z} p_k (k-1) C\left( \frac{1}{k}\right) > 1\,.
\end{equation}
Given the network's degree distribution $p_k$ and the CDF $C$ of thresholds, it is easy to evaluate the condition (\ref{GP12}), so Eq.~(\ref{GP12}) is a very useful criterion for determining whether global cascades can exist (the ``supercritical regime'') or not (the ``subcritical regime'').


\section{Networks with degree--degree correlations}\label{sec:GP3}


We now follow \cite{Gleeson08,Dodds09,Payne09} and extend the message-passing approach to networks with nontrivial degree--degree correlations. Let $p_{k k'}$ be the joint probability distribution function (PDF) for the degrees $k$ and $k'$ of end nodes of a uniformly randomly chosen edge of a network\footnote{In configuration-model networks, in which no correlations are imposed in the generative model, $p_{k k'} = k p_k k' p_{k'}/z^2$, because the degree of nodes at the two ends of an edge are independent.}. As in Sec.~\ref{sec:GP2}, and referring again to Fig.~\ref{fig:GP2}, we define $q_{n}^{(k)}$ as the probability that a degree-$k$ node on level $n$ is active, conditional on its parent (on level $n+1$) being inactive. Similarly, writing $\overline{q}_n^{(k)}$ for the probability that a child of an inactive level-$(n+1)$ node of degree $k$ is active, it follows that
\begin{equation} \label{GP13}
	\overline{q}_n^{(k)} = \frac{\sum_{k'} p_{k k'} q_n^{(k')}}{\sum_{k'} p_{k k'}}\,,
\end{equation}
because a neighbor of the degree-$k$ node has degree $k'$ with probability $p_{k k'}/\sum_{k''}p_{k k''}$. Similar to Eq.~(\ref{GP7}), we then determine the conditional probabilities for each degree at level $n$ from the children at level $n-1$ using the relation
\begin{equation}
	q_n^{(k)} = \rho_0 + (1-\rho_0) \sum_{m=0}^{k-1} {{k-1}\choose m}   \left(\overline{q}_{n-1}^{(k)}\right)^m \left(1-\overline{q}_{n-1}^{(k)}\right)^{k-1-m} C\left(\frac{m}{k}\right)\,, \label{GP14}
\end{equation}
where $q_0^{(k)} = \rho_0$ for all $k$. The unconditional density of active degree-$k$ nodes in steady-state is
\begin{equation}
	\rho_\infty^{(k)} = \rho_0 + (1-\rho_0) \sum_{m=0}^{k} {{k}\choose m}   \left(q_{\infty}^{(k)}\right)^m \left(1-q_{\infty}^{(k)}\right)^{k-m} C\left(\frac{m}{k}\right)\,, \label{GP15}
\end{equation}
and the total network density is equal to
\begin{equation}
	\rho_\infty = \sum_k p_k \rho_\infty^{(k)}\,. \label{GP16}
\end{equation}


\subsection{Matrix criticality condition}

As in Sec.~\ref{sec:GPcc}, one can derive the condition that determines whether global cascades arise from infinitesimal (i.e., $\rho_0\to 0$ seeds) by linearizing the system of equations (\ref{GP14}) about the
zero-contagion solution $q_{n}^{(k)}\equiv 0$ for all $n$ and $k$. Note that Eqs.~(\ref{GP14}) includes one equation for each distinct degree class in a network, so the condition for instability of the contagionless solution is an eigenvalue condition on the Jacobian matrix of the system. From Eqs.~(\ref{GP14}) and (\ref{GP13}), we find (see \cite{Gleeson08}) that the condition for instability (i.e., for the existence of global cascades) is that the largest eigenvalue\footnote{The matrix $\mathbf{M}$ is not symmetric, but there exists a similarity transformation to a symmetric matrix, so all of its eigenvalues are real.} of the matrix $\mathbf{M}$ exceeds $1$, where $\mathbf{M}$ is the matrix with entries
\begin{equation} \label{GP17}
	M_{k k'} = \frac{(k'-1)}{\sum_{k''} p_{k k''}}p_{k k'} C\left(\frac{1}{k'}\right)\,.
\end{equation}
As noted in \cite{Gleeson08}, a similar condition occur for bond percolation on degree--degree correlated networks \cite{Newman02}, and such conditions are also relevant for epidemic models on networks \cite{Kissbook}.


The message-passing method that we have described has also been generalized for networks with community
structure \cite{Gleeson08} and different degree--degree correlations in different communities \cite{MelnikChaos} (where the latter case also has a notable interpretation in the language of multilayer networks \cite{Kivela14}), multiplex networks \cite{Yagan12}, other contagion models \cite{Karsai14}, dynamics in which nodes can be in more than two states \cite{MelnikMulti}, and more. Reference \cite{GleesonPRX13} presented an alternative derivation (starting from the so-called ``approximate master equation'' (AME) framework) of the configuration-model approximation equations (\ref{GP8}) and (\ref{GP10}).


\section{Message-passing for finite-size networks}\label{sec:GP4}

In this section, we discuss message-passing approaches \cite{shrestha14,Lokhov15} that are applicable to finite-size networks, rather than to the ensembles of (infinite-size) networks that we discussed above. Recent papers \cite{shrestha14,Lokhov15} have shown how a message-passing approach can be applied successfully to networks with a finite number of nodes. In this section, we explain this idea by applying it to the WTM. The resulting equations are computationally very expensive to solve. We close the chapter by deriving the analog of the criticality conditions of Eqs.~(\ref{GP12}) and (\ref{GP17}) for the existence of global cascades in finite-size networks. This criticality condition is relatively tractable to compute.



Suppose that we are given an finite-size network that is unweighted and undirected (and unipartite).
The total number of edges in the $N$-node network is $E$, where $E=N z/2$ and $z$ is the mean degree. To use message-passing approach, we consider quantities like $q_{j\to i}$, which are specific for a \emph{directed} edge $j\to i$. We consider each undirected edge of the network (such as the one between nodes $i$ and $j$) as consisting of a reciprocal pair of directed edges ($i\to j$ and $j\to i$), giving a total of $2 E$ directed edges. The direction of the edges gives the local directionality of a contagion, analogous to the ascending levels in Fig.~\ref{fig:GP2}.


\begin{figure}
\centering
\vspace{1cm}
\epsfig{figure=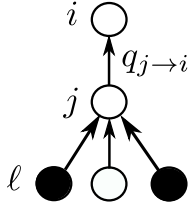,height=2.7cm}
\caption{Schematic for the message-passing approach of Sec.~\ref{sec:GP4}.} \label{fig:GP3}
\end{figure}

The edge-based quantity $q_{j\to i}$ is the probability that node $j$ is active, conditional on node $i$ being inactive. See Fig.~\ref{fig:GP3}, and compare it to Fig.~\ref{fig:GP2}. To write an equation for $q_{j \to i}$, we consider the effect on $j$ of all of its neighbors aside from $i$. Specifically, if node $j$ is not a seed node (which has probability $1-\rho_0$), it is active only if sufficiently many of its neighbors are active. To calculate $q_{j\to i}$, we assume that node $i$ is inactive\footnote{This assumption has various names: it is called the ``cavity approach'' in statistical physics \cite{Mezardbook,zdeborova16,Rogers15}, and it is closely related to the WOR (``without regarding'') property used for financial contagion cascades in \cite{Hurdbook}.}, so we must consider whether the number of active nodes among the remaining neighbors is sufficient to active node $j$.
It is convenient to introduce the notation $\sigma_\ell$ to represent the state of node $\ell$ in a given realization: $\sigma_\ell=1$ if node $\ell$ is active, and $\sigma_\ell=0$ if node $\ell$ is inactive. One can then write the equation for $q_{j\to i}$ as
\begin{equation} \label{GP18}
	q_{j\to i} = \rho_0 + (1-\rho_0)\sum_{ \left\{ \sigma_\ell \right\}: \ell \,\in {\mathcal{N}_j}\setminus i} {C\left(\frac{\sum \sigma_\ell}{k_j}\right) \prod_{\sigma_\ell=1} q_{\ell\to j} \prod_{\sigma_\ell=0}\left(1-q_{\ell\to j}\right)}\,.
\end{equation}
The summation in Eq.~\ref{GP18} is over all combinations of $\sigma_\ell$ values. In other words, one sums over the possible states of the neighbors of $j$ (where $\mathcal{N}_j$ denotes the set of such neighbors), except for node $i$. Given the set of neighbor states $\left\{ \sigma_\ell \right\}$, the fraction of active neighbors of node $j$ is ${\sum \sigma_\ell}/{k_j}$, where ${k_j}$ is the degree of node $j$, and the probability that this fraction is at least as large as the threshold of node $j$ is given by $C\left(\frac{\sum \sigma_\ell}{k_j}\right)$.
Let's consider each of the inactive node $j$'s neighbors, except for $i$. Because each of these nodes $\ell$ is active with an independent probability of $q_{\ell \to j}$, the first product term of Eq.~\eqref{GP18} gives the probability that a specified subset of nodes is active, and the second product term of Eq.~\eqref{GP18} gives the probability that the remaining neighbors of $j$ are inactive.
Consequently, multiplying the two product terms gives the probability (assuming $j$ is inactive) to have a given combination $\left\{ \sigma_\ell \right\}_{\ell \,\in {\mathcal{N}_j}\setminus i} $ of neighbors' states, and the sum over all possible combinations plays the same role as the sum over $m$ in Eqs.~(\ref{GP7}) and (\ref{GP14}).

In principle, one can solve Eq.~(\ref{GP18}) by iteration to determine $q_{j \to i}$ for every directed edge. The probability that node $i$ is active (similar to Eq.~(\ref{GP15})) is then given by
\begin{equation} \label{GP19}
	\rho^{(i)} = \rho_0 + (1-\rho_0) \sum_{ \left\{ \sigma_j \right\}: j \,\in {\mathcal{N}_i}} {C\left(\frac{\sum \sigma_j}{k_i}\right) \prod_{\sigma_j=1} q_{j\to i} \prod_{\sigma_j=0}\left(1-q_{j\to i}\right)}\,,
\end{equation}
where the sum in Eq.~\eqref{GP19} is over all neighbors of $i$ (compare to Eq.~(\ref{GP15})). Unfortunately, the summations in both Eqs.~(\ref{GP18}) and (\ref{GP19}) require calculating a combinatorially large numbers of terms. For example, the sum over the sets $\left\{ \sigma_\ell \right\}_{\ell \,\in {\mathcal{N}_j}\setminus i} $ of the possible states of the neighbors of node $j$ has $2^{k_j-1}$ terms, each of which has its own probability measure that needs to be evaluated with the two product terms in Eq.~(\ref{GP18}).  The large number of possible combinations makes the implementation of this message-passing approach extremely computationally expensive, except for very small networks.  


On the bright side, one can derive the steady-state equations for the configuration-model ensemble that we discussed in Sec.~\ref{sec:GP2} from the message-passing equations (\ref{GP18}) and (\ref{GP19}), as is described in detail in \cite{shrestha14}. Essentially, in a configuration-model ensemble, each edge-based conditional probability $q_{\ell \to j}$ is replaced by the single quantity $q$ (which we called $q_\infty$ in Sec.~\ref{sec:GP2}). Because all neighbors are treated as identical, the sum in Eq.~(\ref{GP18}) over
$\left\{ \sigma_\ell \right\}$ becomes the sum over the
number $m$ of active neighbors, weighted by the binomial coefficient ${k-1}\choose m$, which gives the number of arrangements of precisely $m$ active neighbors among the $k-1$ neighbors who could be active. Consequently, the sum over $\left\{ \sigma_\ell \right\}$ in Eq.~(\ref{GP18}) reduces to a sum over $m$ in Eq.~(\ref{GP6}), yielding the steady-state limit ($n\to \infty$) of the configuration-model equations (\ref{GP8}) and (\ref{GP10}).



\subsection{Criticality condition for finite-size networks}

Although calculating the full message-passing equations \eqref{GP19} is prohibitively expensive for large networks, one can nevertheless apply the same approach as in earlier sections to derive a condition for the existence of global cascades.
As before, we take the $\rho_0 \to 0$ limit and linearize the governing equation (\ref{GP18}) about the zero-contagion equilibrium.  Specifically, we linearize Eq.~(\ref{GP18}) about $q_{j\to i}=0$ for each edge. For very small values of the edge probabilities, the sum in Eq.~(\ref{GP18}) gives a linear contribution only when a single neighbor is active. The resulting linearization is then given by
\begin{equation}
	q_{j \to i} = \sum_{\ell \in \mathcal{N}_j \setminus i} C\left( \frac{1}{k_j}\right) B_{i\to j, j\to \ell} \,\, q_{\ell\to j} \label{GP20}\,,
\end{equation}
where $\mathbf{B}$ is the nonbacktracking (Hashimoto) matrix, which has recently been studied in network-science questions such as percolation \cite{Karrer14}, community detection \cite{MoorePNAS14}, and centrality \cite{MartinNewman, Radicchi16a}. The nonbacktracking matrix is a sparse matrix of dimension $2E \times 2 E$, where each row (or column) corresponds to a directed edge between two nodes. The elements of $\mathbf{B}$ are nonzero when the directed edge corresponding to the row (e.g., the edge $i\to j$) leads to the directed edge corresponding to the column (e.g., $j\to \ell$) via a common node (which, in this case, is node $j$), provided that the second directed edge does not return to the source node of the original edge (i.e., node $\ell$ cannot be the same as node $i$).

Rewriting Eq.~(\ref{GP20}) in a matrix form that is suitable for iteration
(analogous to Eqs.~(\ref{GP8}) and ~(\ref{GP14})) yields
\begin{equation}
	\mathbf{q}_{n} = \mathbf{D} \mathbf{B}\mathbf{q}_{n-1}\,, \label{GPDB}
\end{equation}
where $\mathbf{q}$ is the $2E$-vector of values $q_{j\to i}$. We then immediately see that the linear stability of the $\mathbf{q}=\mathbf{0}$ solution depends on the largest eigenvalue of the product matrix $\mathbf{D}\mathbf{B}$, where $\mathbf{D}$ is a $2E\times 2E$ diagonal matrix with nonzero elements given by
\begin{equation}
	D_{i\to j,i\to j}= C\left(\frac{1}{k_j}\right)\,.
\end{equation}
The criterion that we have derived from the message-passing approach is therefore that the existence of global cascades requires the spectral radius of the $2E\times 2E$ matrix $\mathbf{D}\mathbf{B}$ to exceed 1. Since the matrix is sparse, this cascade criterion can be checked even for large networks.

\begin{table}
\begin{center}
\begin{tabular}{|c|c|c|c|c|}
\hline
Network & $N$ & $z$ & $\theta_\text{config}$ & $\theta_\text{crit}$ \\
\hline\hline
3-regular random graph & $10^5$ & 3 & $\frac{2}{3}$ & $\frac{2}{3}$ \\\hline
Facebook Caltech \cite{Traud11} & $762$ & 43.7 & $ 0.98 $ & $ 0.98 $ \\\hline
Facebook Oklahoma \cite{Traud11} & $17420$ & 102 & $ 0.99 $ & $ 0.99 $ \\\hline
Gowalla \cite{SNAPGowalla,SNAPdataGowalla} & $1.97\times10^5$ & 9.67 & $ 0.90 $ & $ 0.94 $ \\\hline
PGP network \cite{PGP2,PGP3} & $10680$ & 4.55 & $ 0.78 $ & $ 0.94 $ \\\hline
Power grid \cite{powergrid1,powergrid2} & $4941$ & 2.67 & $ 0.63 $ & $ 0.78 $ \\\hline
\hline
\end{tabular}
\caption{The critical value of $\theta$, the upper limit of the uniform distribution of thresholds, for the WTM on various networks, as calculated using the configuration-model result Eq.~(\ref{GP12}) for $\theta_\text{config}$ and using the maximum eigenvalue of the $\mathbf{D} \mathbf{B}$ matrix in Eq.~(\ref{GPDB}) to determine $\theta_\text{crit}$. The network size (i.e., number of nodes) is $N$ and the mean degree is $z$, so the number of undirected edges is $E=N z/2$. Note, as expected, that $\theta_\text{config}$ is identical to $\theta_\text{crit}$ for the 3-regular random graph. The corresponding values for the Facebook networks are also very close, indicating that the configuration-model theory is very accurate for these networks (as also found in \cite{Melnik11,Gleeson12}). For the other networks, there is a considerable difference between $\theta_\text{config}$ and $\theta_\text{crit}$, indicating that the configuration-model result is inaccurate on these networks (although it is also known that the message-passing approach, being based on a tree-like assumption of independence of messages \cite{Faqeeh15}, is inaccurate for spatial networks \cite{Radicchi15,Radicchi16a} such as the power-grid example in this table).} \label{tab:GP1}
\end{center}
\end{table}


In Table~\ref{tab:GP1}, we give examples in which we consider the WTM with thresholds uniformly distributed over the interval $(0,\theta)$, so the mean threshold value is $\theta/2$. If the parameter $\theta$ is small, all thresholds are low, and a seed node is likely to cause many neighbors to become active, leading quickly to a global cascade. However, a very large $\theta$ value implies that many nodes' thresholds are too high to allow them to activate, so no global cascades occur. In Table~\ref{tab:GP1}, we report the critical value of the parameter $\theta$ that separates the global-cascade (i.e., supercritical) regime from the no-global-cascade (i.e., subcritical) regime for several real-world networks using the configuration-model condition given by Eq.~(\ref{GP12}) and the spectral condition on the $\mathbf{D}\mathbf{B}$ matrix that we described above. In previous work on calculating percolation thresholds for real-world networks \cite{Radicchi15,Radicchi15b}, the use of the nonbacktracking matrix has led to more accurate predictions than those found by applying configuration-model theory (which uses only the degree distribution of a network). We therefore anticipate that the cascade threshold identified by the largest eigenvalue of the $\mathbf{D}\mathbf{B}$ matrix will prove to be more accurate than configuration-model predictions and will shed further light on the structural features of certain networks that enable configuration-model theories to give accurate results \cite{Gleeson12}.


\section{Conclusions}\label{sec:GP5}


In this chapter, we reviewed several analytical approaches for complex-contagion dynamics. For concreteness, we focused on the example of the Watts threshold model, but the methods that we discussed can also be applied to other monotonic binary-state dynamics \cite{GleesonPRX13}.
 To provide context, we first introduced a naive mean-field approach, which has limited accuracy. We then showed that using the methods of \cite{Dhar97,Gleeson07} gives very accurate results on configuration-model networks. We demonstrated how the methodology can yield a criterion for determining whether global cascades occur, and we briefly reviewed an extension of the method to networks with imposed degree--degree correlations. In Sec.~\ref{sec:GP4}, we briefly discussed the approaches of \cite{shrestha14,Lokhov15} to derive message-passing equations for cascades on finite-size networks. Although the resulting equations are computationally expensive to solve, we showed that they give a condition for global cascades in terms of the spectral radius of a matrix that is related to the nonbacktracking matrix. The nonbacktracking matrix has arisen in prior work from linearizations of belief-propagation algorithms \cite{MoorePNAS14}, but the product matrix $\mathbf{D}\mathbf{B}$ that determines the cascade condition has not been studied in detail (to our knowledge), and we believe that further investigations of it will yield fascinating insights into the propagation of complex contagions and other monotonic dynamics \cite{Lokhov15,Lokhov16}.



\section*{Acknowledgements}

This work was supported by Science Foundation Ireland (grant numbers 15/SPP/E3125 and 11/PI/1026). We
acknowledge the SFI/HEA Irish Centre for High-End Computing (ICHEC) for the provision of
computational facilities. 



\bibliographystyle{unsrt}
\bibliography{message_passing_refs}

\end{document}